\documentclass[iop]{emulateapj}		
\usepackage{epsfig}			
\usepackage{graphicx,color}		
\usepackage{amssymb}			
\usepackage{color}			
\usepackage{url}			
\usepackage{amsmath}			
\usepackage{rotating}			
\usepackage{float}			
\usepackage{textcomp}			
\usepackage{psfig}
\usepackage{epstopdf}
\usepackage{dcolumn}
\usepackage{times}
\usepackage{tabularx}
\usepackage[english]{babel}
\usepackage[normalem]{ulem}
\usepackage{float}
\usepackage{ragged2e}

\shorttitle{Supergranule Scale and TSI}
\shortauthors{Sudip Mandal et al.}

\begin{document}


\title{
Association Of Supergranule Mean Scales with Solar Cycle Strengths and Total Solar Irradiance
}

\author{Sudip Mandal $^{1}$,
        Subhamoy Chatterjee$^{1}$,
        Dipankar Banerjee$^{1,2}$
}
 
\affil{$^{1}$Indian Institute of Astrophysics, Koramangala, Bangalore 560034, India. e-mail: {\color{blue}{sudip@iiap.res.in}}\\
$^{2}$ Center of Excellence in Space Sciences India, IISER Kolkata, Mohanpur 741246, West Bengal, India  \\}

\begin{abstract}
{ We analyse the long-term behaviour of supergranule scale parameter, in active and quiet regions (AR, QR), using the Kodaikanal digitized data archive. This database provides century-long daily full disc observations of the Sun in  C\MakeLowercase{a} $\scriptsize{{\textrm{II}}}$ K wavelength. In this paper, we study the distributions of the supergranular scales, over the whole data duration, which show identical shape in these two regimes. We found that the AR mean scale values are always higher than that of the QR for every solar cycle. The mean scale values are highly correlated with the sunspot number cycle amplitude and also with total solar irradiance (TSI) variations. Such correlation establishes the cycle-wise mean scale as a potential calibrator for the historical data reconstructions. We also see an upward trend in the mean scales, as already been reported in TSI. This may provide new input for climate forcing models. These results also give us insight into the different evolutionary scenarios of the supergranules in the presence of strong (AR) and weak (QR) magnetic fields.}
\end{abstract}

\section{Introduction}

  Solar cycle refers to the 11 year periodic variation observed in a variety of solar activities such as sunspot number, sunspot area, 10.7 $\mathrm{cm}$ radio flux, flare occurrences, number of coronal mass ejections (CMEs) etc (see \citet{lrsp-2015-4} for a complete review on solar cycle). Almost all the phenomena/features on the Sun, seem to have this periodicity embedded into their properties. In this paper, we study about a chromospheric feature, known as supergranules. These are the large scale (~25-30 $\mathrm{Mm}$) velocity ($\sim$ 400 $\mathrm{m/sec}$) structures on the solar surface, with an average lifetime of 25 hours \citep{Rieutord2010}.

Study of supergranules from the ground based telescopes has been an interest since early 1900's. Though there are reports as early as 1916 \citep{1916ApJ....43..145P}, the first confirmation came from \citet{1956MNRAS.116...38H} who reported on `velocity fluctuations' having a length scale of 26 $\mathrm{Mm}$. In the next few decades, a lot of work has been done on supergranules including the first confirmation from Doppler images by \citet{1962ApJ...135..474L}. \citet{1964ApJ...140.1120S} realized that the strong magnetic network-like structures (known as `network boundaries'), seen through chromospheric C\MakeLowercase{a} $\scriptsize{{\textrm{II}}}$ K line, are actually the tracer of supergranule cells. Later, \citet{0004-637X-597-2-1200} showed that the evolution of the supergranules trigger the dispersion of the magnetic field stored in these boundaries.

Thus, the first step in analysing the supergranular properties, is to detect these network boundaries from C\MakeLowercase{a} $\scriptsize{{\textrm{II}}}$ K images. Several techniques have been applied by many authors to detect the network pattern (hence supergranules): auto-correlation technique \citep{1970SoPh...13..292S,0004-637X-481-2-988}, Fast Fourier Transformations (FFT)  \citep{1989A&A...213..431M}, automated skeleton detection \citep{1999A&A...344..965B}, watershed transformation \citep{2041-8205-730-1-L3,2017arXiv170500175C}. Several properties of the supergranules, such as scale \citep{0004-637X-534-2-1008,2041-8205-730-1-L3}, fractal dimension \citep{Paniveni11022010} shows signatures of a solar cycle like variation. In fact, \citet{0004-637X-481-2-988,1998SoPh..180...29B} showed that the scale distribution follow the voronoi tessellation pattern.  Most of these results were derived using a short span of data (less than a solar cycle) or a couple of solar cycles \citep{2041-8205-730-1-L3}.

The other important connection of supergranule scale was found by \citet{2041-8205-730-1-L3} where the authors showed an in-phase variation of the scale parameter, for the 23$^{rd}$ cycle, with the `Total Solar Irradiance' (TSI) cycle.
Historically, this TSI was thought to be constant and considered only to be variable in longer time period (evolutionary time scale of the star). Only after the advent of space borne radiometers, it became clear that the TSI varies in much smaller time scales, from days to years \citep{1981Sci...211..700W}. Since space based measurements are only available from 1970's, there are studies which link the historical sunspot observations with the TSI changes on solar cycle time scales \citep{1982SoPh...76..211H,1998GeoRL..25.4377F,2004A&ARv..12..273F}.
Thus, a long-term study of the scale property will enable us to better understand the historical TSI variations. 

One of the key constraints in establishing a clear connection between the different supergranular properties with the solar cycles, is the non-availability of a long-term uniform, homogeneous dataset. Kodaikanal Solar Observatory (KSO) in India, has been accruing C\MakeLowercase{a} $\scriptsize{{\textrm{II}}}$ K observations since 1907. For the first time, these images have been used by \citet{2017arXiv170500175C} to detect the supergranules in an automated manner and derived different properties of them over the span of 100 years (cycle 14-23). These authors found that the supergranule scales and fractal dimension have the 11 year solar cycle like periodicities persistent throughout the data. One of the highlight of this work, is the opposite correlations of the active and quite region supergranule scales with the sunspot number cycle.

As a follow up  of \citet{2017arXiv170500175C} we explore further on the properties of supergranule scale. Section~\ref{sec:inter_cycle} presents investigation of supergranule mean scale variation over two regions. Long term association of these scale values with the TSI and the sunspot number has been analyzed in Section~\ref{sec:correlations}. Next, the variation of the scale parameter within a given cycle is derived in Section~\ref{sec:intra_cycle}. We conclude by summarizing the main results in Section~\ref{sec:summary}. 

\section{Kodaikanal Data Description} \label{sec:data}
We have used the digitized archive of C\MakeLowercase{a}  $\scriptsize{{\textrm{II}}}$ K observations captured at Kodaikanal Solar observatory, India. Originally the images have been taken in photographic plates and recently, these plates have been digitized in a 4K$\times$4K format at Indian Institute of Astrophysics. Digitized version of this data, is now available for public use at \url{https://kso.iiap.res.in//}. This data cover a period of almost 104 years i.e from cycle 14 (decline phase only) to cycle 23. Since there are significant number of missing days and poor plate conditions after 1996 \citep{2014SoPh..289..137P,2016ApJ...827...87C}, we only consider cycle 14 to cycle 22 in this study.

\section{Method} \label{sec:method}

 We use the detected network boundary information as obtained by \citet{2017arXiv170500175C}. This feature extraction was done using an automated algorithm based on the `watershed technique' \citep {CYTO:CYTO10079,2041-8205-730-1-L3}. We again remind the reader that these network boundaries are the proxies for the supergranular cells \citep{1964ApJ...140.1120S}. Regions on the sun has been divided into two regimes: active regions (ARs) and quiet regions (QRs) (see \citet{2017arXiv170500175C} for more details). In this study, we use the scale parameter for each of the supergranules from daily images. The supergranule scale is defined as the radius of the circle whose area is equal to the area of the supergranule \citep{2017arXiv170500175C}.

\section{Results}

\subsection{Inter-cycle Variations}\label{sec:inter_cycle}

\begin{figure}[!htb]
\centering
\includegraphics[width=0.48\textwidth]{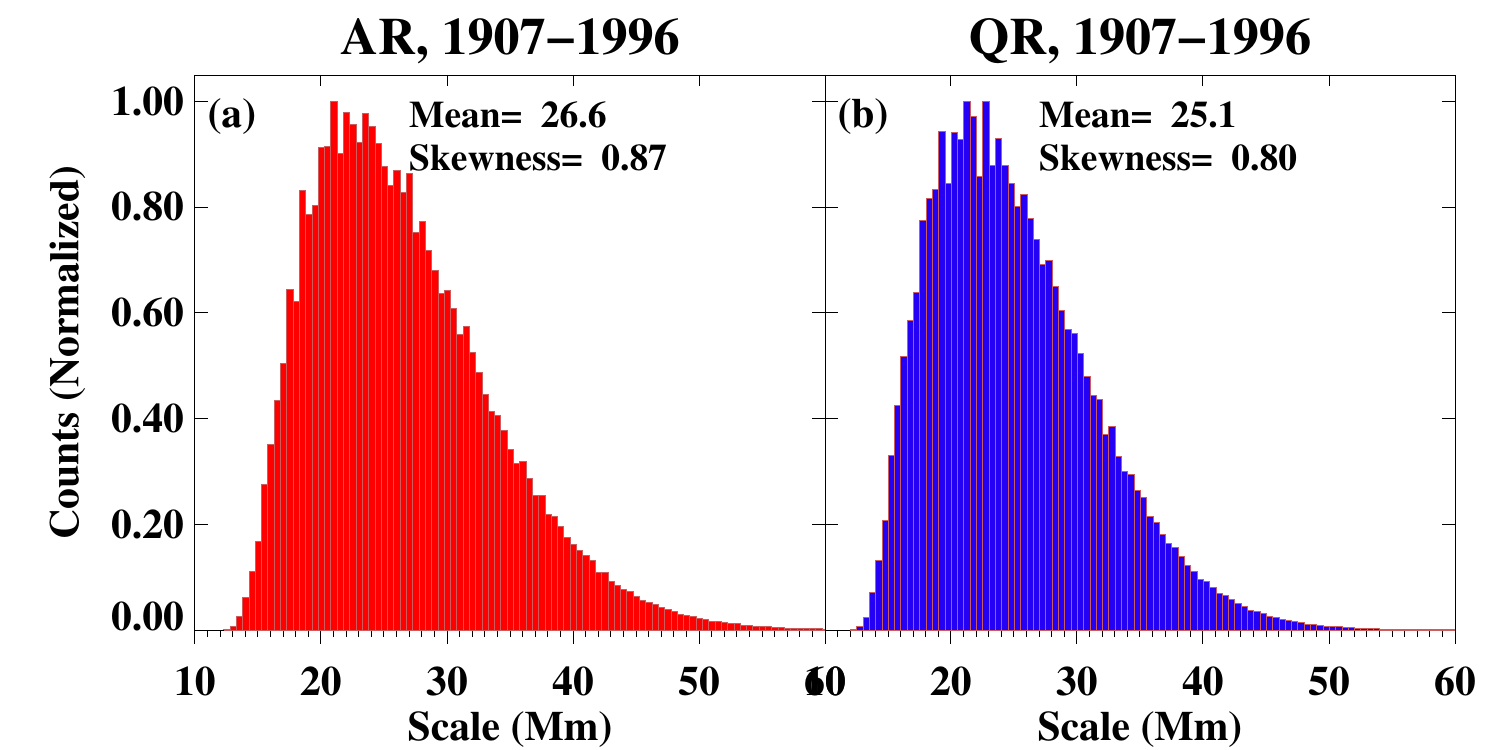}
\caption{Distributions of the supergranule scales for the AR (panel a) and the QR (panel b) respectively. The histograms are constructed from daily observations over the period of 1907 to 1996.}
\label{hist}
\end{figure}

First we compare the distribution of the supergranular scale values obtained from daily images over the whole span of the data (cycle 14-22), separately for AR and QR. The normalized histograms are plotted, in panels (a-b) in Figure~\ref{hist} (red for AR and blue for QR). From the plot we readily notice that the distributions in two regimes are of similar shape (note the closeness of the skewness parameter, 0.87 for AR and 0.80 for QR). This is interesting, since the AR and QR represent different magnetic field strengths and it has been conjectured by \citet{2008A&A...488.1109M} that the network magnetic elements have a shrinking effect on the supergranules. Also, both of these distributions resemble well with the voronoi distribution \citep{0004-637X-481-2-988}. In other-words, such resemblance also indicate the goodness of the automated detection algorithm used by \citet{2017arXiv170500175C}. We also note that the mean of this scale distribution in the AR is higher (26.6 $\mathrm{Mm}$) than that of the QR (25.1 $\mathrm{Mm}$). This is to remind that these values are calculated over 9 solar cycles. This means, it is not clearly evident whether such differences in the scales are persistent over all the cycles or just an anomalous cycle lead to this overall contrast.\\

We probe such scenarios by calculating the mean scale for every individual cycle, for the AR as well as for the QR.
 While deriving the mean values, we also calculated the uncertainties involved in those values. One of the primary sources of uncertainties is the watershed detection method itself. In fact, \citet{2041-8205-730-1-L3}, who have used this technique first to detect the supergranules, have also performed the error measurements involved in this technique by utilizing 1000 Monte-­Carlo realizations.
Using a Poisson noise model, these authors found that the changes in the mean values are considerably less than the annual and monthly variations in the value itself (they found the variance of the mean value to be $\sim$0.01Mm). Thus, we derive the errors of the average values by calculating the `confidence interval' $\mathrm{(C.I)}$. Since the sample standard  deviation is known, so the formula for the $\mathrm{(C.I)}$ within which population mean should exist, reads as:

\begin{equation}
  \mathrm{C.I}= \bar{X} + Z^{*} \times (\mathrm{s}/\sqrt{\mathrm{n}})
\end{equation}

 where $\bar{X}$ is the sample mean value whereas $\mathrm{s}$ and $\mathrm{n}$ are the standard deviation and the sample size  respectively. $\mathrm{Z}$$^{*}$ is the multiplier value which is set to 1.96 which corresponds to  a confidence interval of 95\%. Now, this number is derived from a normal distribution. From Figure~\ref{hist}, we note that the histograms are not normally distributed (in-fact positively skewed). Since our sample size is very large ($\approx$60,000), we make use of the same multiplier value according to the `central-limit theorem'.\\

\begin{figure*}[!htb]
\centering
\includegraphics[width=0.95\textwidth]{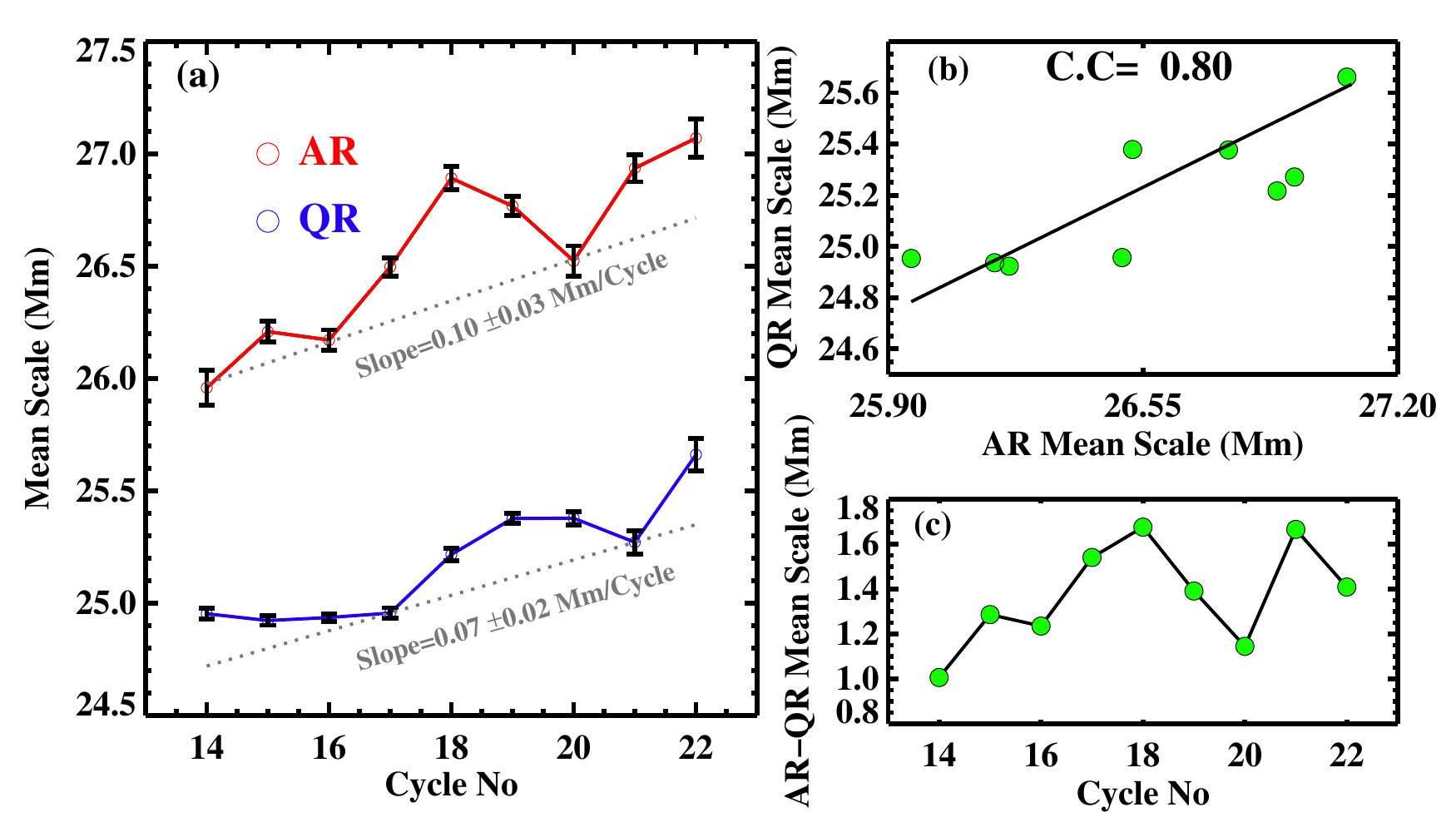}
\caption{ Panel (a) shows the variations of the mean supergranular scales with the solar cycles for AR and QR. The scatter plot between the two, is shown in panel (b) whereas differences in the AR and QR mean scales are plotted in panel (c). The error bars in panel (a) represent the 95\% confidence limit on the calculated mean scale values.}
\label{inter_cycle}
\end{figure*}

 The results are plotted in panel (a) of Figure~\ref{inter_cycle}. There are quite a few interesting features visible in this plot. Firstly, we notice that the AR mean scale is always higher than that of the QR, for all the cycles. Thus we confirm the persistence as questioned in the previous section. Next, we see that both the AR and QR mean scales are oscillating with a common pattern as we go through different cycles. A careful inspection, reveal that such variation patterns are also present in sunspot cycles which we will explore in details in following sections.

We again look back at these curves to explore more features from them. Both the curves, seem to ride on an upward moving trend. This feature is particularly interesting as such trends are not common in solar proxies but found in TSI from observations \citep{1997Sci...277.1963W}. Though within our data duration, we only have two distinct minima points, we try to quantify the trends by fitting the minimum points with straight lines (dotted lines in panel~\ref{inter_cycle}.a). In order to estimate these slopes along with the uncertainties, we use the confidence interval of the mean values as the measurement errors during the linear fitting. For the AR case the slope value is 0.10 $\pm$ 0.03 $\mathrm{Mm}$/year whereas for the QR, it is 0.07 $\pm$ 0.02 $\mathrm{Mm}$/year. Thus the trend is marginally steeper in AR compared to QR. Though these slope values are small, but on a longer time scales (for example, on decadal time scales) they have a profound effect on the mean scale values.

Though the two curves show a similar variation with cycles, but there are some subtle differences also. The locations of the maxima and minima of the two curves appear at different cycle numbers. For an example, the maxima for the AR mean scale occurs for the 18$^{th}$ cycle whereas the QR curve reaches it maximum at the 19$^{th}$ cycle (same happens for the minimum also). Such small deviations in the two curves, result in a correlation value of 0.82 between them as shown in panel~\ref{inter_cycle}.b. Though we have already seen the dominance of AR mean scales over the QR values, but it is necessary to quantify the difference between the two. In panel~\ref{inter_cycle}.c, we plot the scale difference (AR$-$QR) with different solar cycles. Interestingly, this difference value reaches to its maximum value of 1.7 $\mathrm{Mm}$ twice (18$^{th}$ and 21$^{th}$ cycle) whereas the minimum value of 1 $\mathrm{Mm}$) also occurs twice (14$^{th}$ and 20$^{th}$ cycle).

\subsection{Relations with Sunspot Number and TSI}\label{sec:correlations}

We have seen in the previous section that the AR and QR mean scales show oscillatory patterns (with solar cycles) which show resemble to the amplitude variation of sunspot cycles. In fact, as mentioned earlier, \citet{2041-8205-730-1-L3} have shown the association of the scale values with the TSI variation only for the 23$^{rd}$ cycle. In this section, we extend the study using the data over 9 solar cycles and build some statistics.

\begin{figure}[!htb]
\centering
\includegraphics[width=0.48\textwidth]{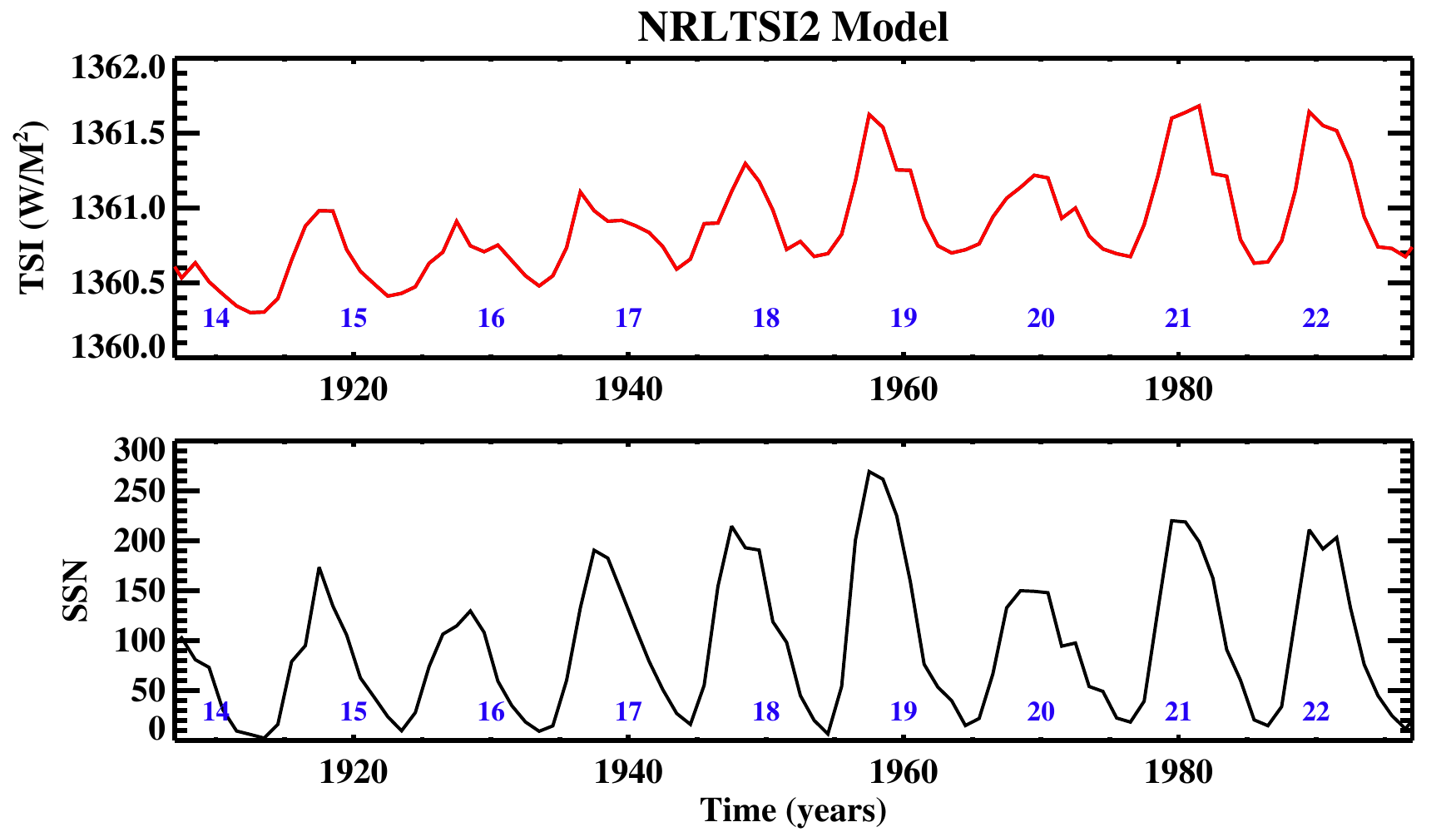}
\caption{Top panel shows the yearly variation of the TSI obtained from NRLTSI2 model. The bottom panel shows the same but for the SSN. Solar cycle numbers are also printed near the bottom of the panels.}
\label{tsi_context}
\end{figure}

The earliest TSI observations are available only from 1980's from Active Cavity Radiometer Irradiance Monitor (ACRIM 1,2,3) followed by Variability of solar IRradiance and Gravity Oscillations (SOHO/VIRGO) (since 1996) and Total Irradiance Monitor (SORCE/TIM) (since 2003). Within the data period investigated in this paper, we have only one and a half solar cycle data overlap. Thus, we make use of the historical TSI reconstruction model output. The NRLTSI2 (Naval Research Laboratory's Total Solar Irradiance) historical TSI reconstruction (for the last 400 years) is based on the model by \citet{1995GeoRL..22.3195L}. The NRLTSI data are being produced by NOAA as a Climate Data Record and are available at \url{http://lasp.colorado.edu/home/sorce/data/tsi-data/}. The details of the model algorithm and its features are described in \citet{2016BAMS...97.1265C}. This model TSI data has also been updated after the revision of the sunspot number \citep{Kopp2016} and is also scaled to match SORCE/TIM observations.
The model output is plotted in Figure~\ref{tsi_context} along with the yearly SIDC sunspot number data (which is available at \url{http://sidc.oma.be/silso/}).\\

\begin{figure}[!htbp]
\centering
\includegraphics[width=0.48\textwidth]{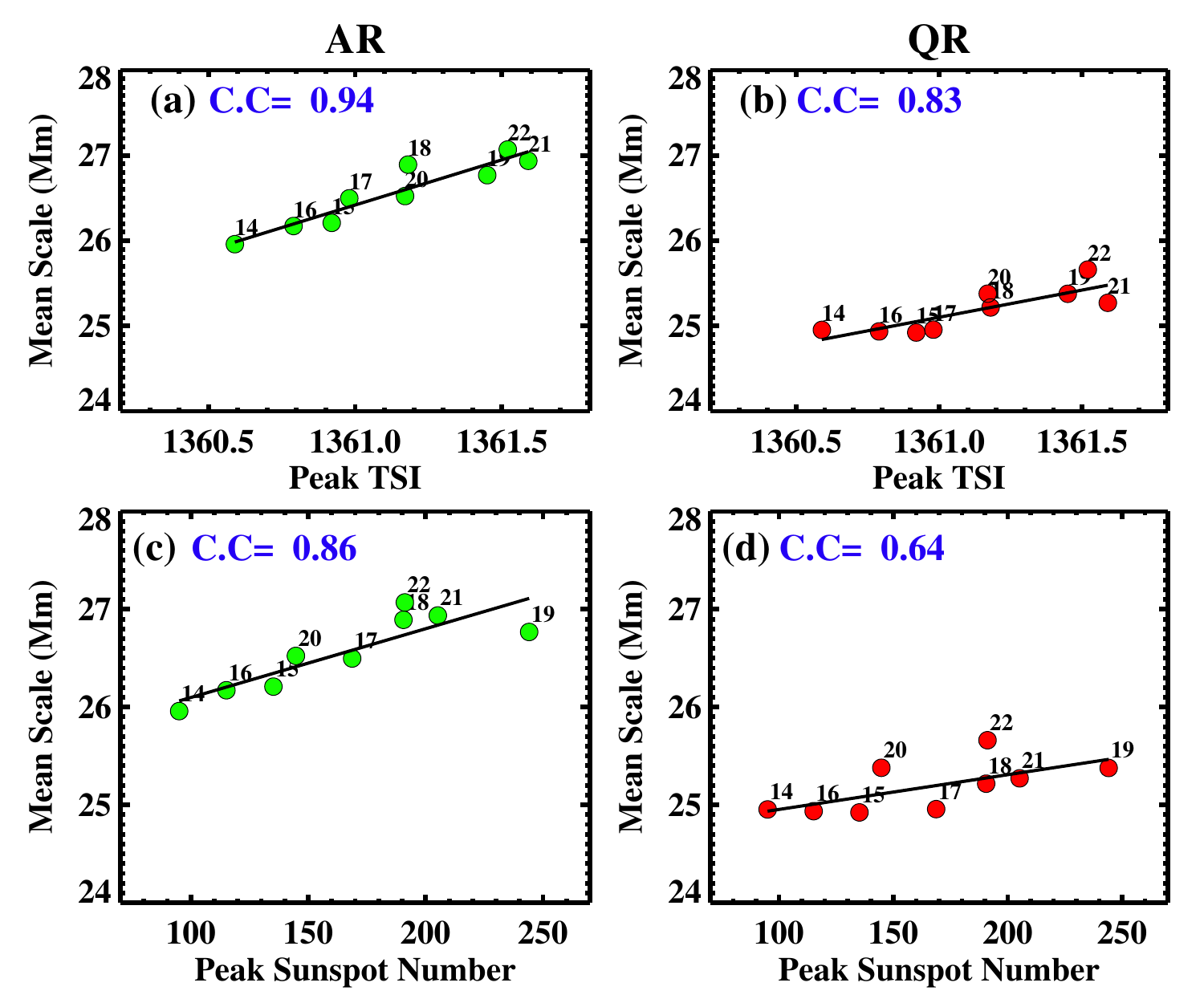}
\caption{Panels (a-b) show the scatter plot between the peak TSI and the mean scale values for different solar cycles whereas panel (c-d) shows the scatter plots between the peak SSN and the mean scale values.}
\label{tsi_corr}
\end{figure}

Immediately from Figure~\ref{tsi_context}, we note a clear resemblance with Panel~\ref{inter_cycle}.a. The TSI also seem to ride on an upward trend with the solar cycles, just like the mean scale curves. In order to have a quantitative measure, we first calculate the peak TSI values (in $\mathrm{W/m^2}$) and peak sunspot number for each of the cycles from Figure~\ref{tsi_context}. In panels (a-b) of Figure~\ref{tsi_corr}, we show the scatter plots between the peak TSI and the mean scale (for both the regions, AR and QR) calculated for each of the cycles. Now, the correlation coefficients (C.C), for the two cases are high but different: for AR it is 0.94 whereas for QR it is 0.83.  Such high correlations, immediately point towards the possible use of the cycle-wise mean scale as a calibrator for the historical TSI data reconstructions. This has another application, as we have seen that the an upward trend is present for both the curves (TSI and scale). Thus one can cross-calibrate between the trend values to set better constraints on it as this trend has an important effect on climate forcing models \citep{lockwood2007recent}.

 We also calculated the correlations between the mean scale and the peak sunspot number of the cycles and are shown in panels (c-d) of Figure~\ref{tsi_corr}. In this case too, we notice that he AR correlation (C.C=0.86) is significantly higher than that of the QR. In order to explain the relatively lesser correlation coefficient for the QR regime, we note that, from Figure~\ref{inter_cycle}, for the initial 4 cycles (cycle 14-17) the QR scale values did not change (constant at $\sim$ 25 $\mathrm{Mm}$) and this may have contributed towards this lower coefficient value. Since there are studies which indicate  small differences between sunspot number and sunspot area series \citep{2016SoPh..291.2917L}, we have thus, computed the correlations between the mean scale and peak sunspot area also and the coefficient values (0.87 for AR and 0.66 for QR) are very similar to the previous values obtained using peak sunspot numbers.

   In an interesting observation, we notice that all the correlation coefficient values computed with peak SSN are less as compared to the same when obtained with peak TSI. This means that the overall supergranular scale is better following the TSI cycle amplitude as compared to the sunspot number cycle amplitude. One possible reason behind this could be the linearly increasing trend which is present for both the TSI and the mean scales but not in the SSN.

\subsection{Histogram Analysis, Intra-cycle Variations}\label{sec:intra_cycle}
\begin{figure*}[!htb]
\centering
\includegraphics[width=0.90\textwidth]{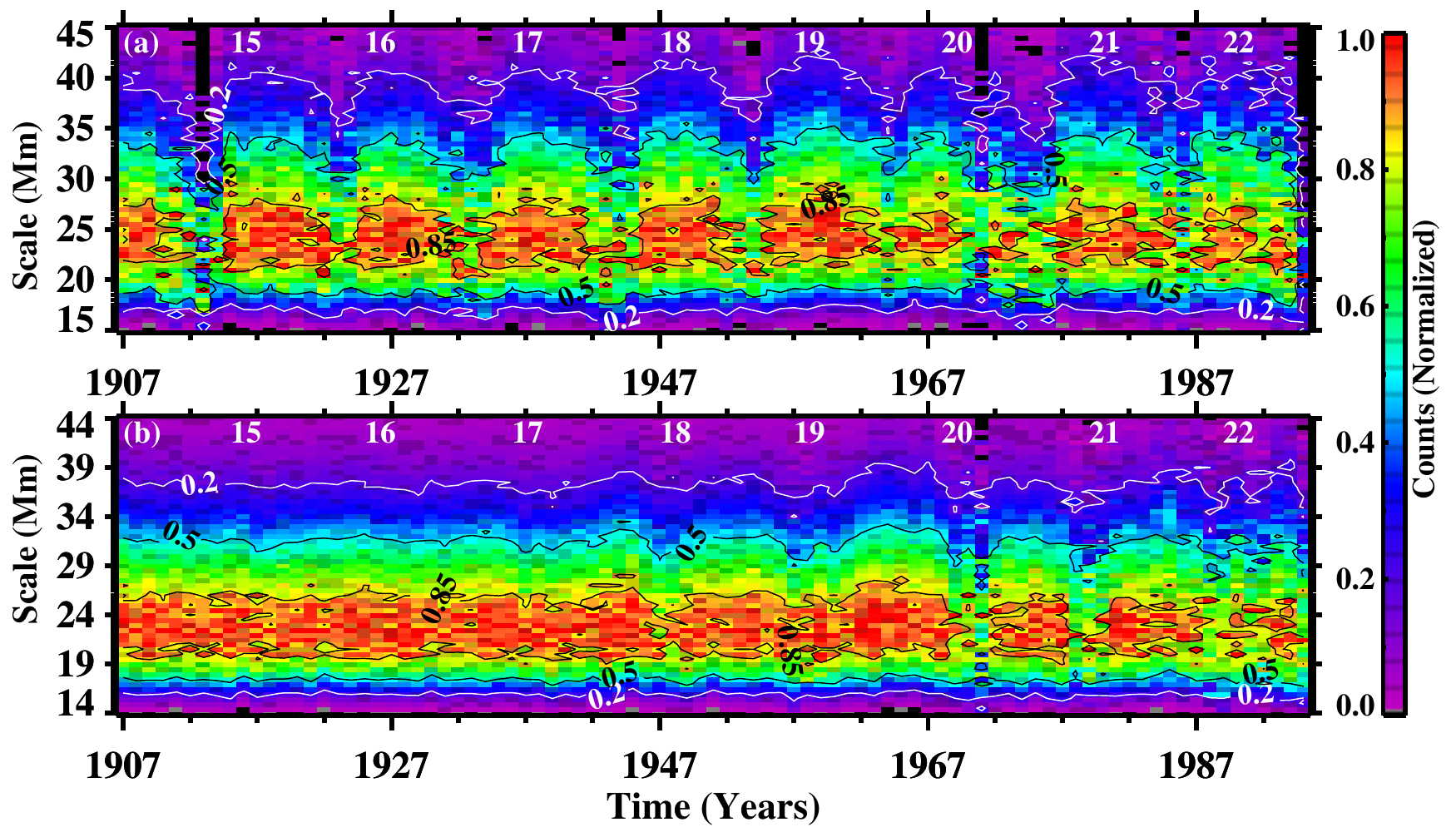}
\caption{2D histograms for the AR (panel a) and QR (panel b) respectively. Different contours indicate the normalized histogram values. See the text for details. }
\label{2d_hist}
\end{figure*}

In the previous two sections, we found that the mean scale values change considerably with different solar cycles and they have an in-phase amplitude variation with the peak sunspot number and the TSI values. On the otherhand, it is expected that the scale distribution will also change within a solar cycle i.e on a time scale of $\approx$11 years. To probe such changes, we construct 2-D histograms for AR and QR and plotted them in panel~\ref{2d_hist}.a and panel~\ref{2d_hist}.b respectively. The y-axes of these plots represent the normalized histograms. The colorbar indicate the strengths of the histograms. In-order to minimize statistical fluctuations, we use yearly data to construct these histograms. Such a 2-D histogram not only captures the dominant scale locations over the time but it allows us to study the other statistical features such as changes in distribution widths (variance). To highlight the features, we use contours of different levels (or strengths): C1 which outlines the higher (0.85) values where the other two contours, C2 and C3, outline the moderate (0.5) and lower (0.2) values respectively.

 We shall discuss the AR histogram first (panel~\ref{2d_hist}.a). In this case, we see that all the three contours show a clear 11 year periodic solar cycle like variations. Now following the C1 contour, for a given cycle, we notice a very interesting property. The contour width is large near the start of a cycle. After that it increases by a small amount and then decreases significantly to reach its minimum value near the end of the cycle. This means that such variation is not symmetric with the cycle duration which is not a common case with the solar proxies (e.g sunspot number or with sunspot area cycle). In another observation, we note that the variations near the wings of the distributions (in C2 and C3) are more prominent compared to the core of the distributions. In this case, these variations follow the shape of the solar cycles which eventually lead to an in-phase alteration of the variance.

Time evolution of the QR histogram (panel b) is different from the AR case though. We note that the width of the C1 contour changes very little with the evolution a cycle. Such a (near) constant width indicates that the the core of the distribution do not change with solar cycle. Though we must again highlight, the effect of the missing data, for the last couple of years, is evident in the last cycle (cycle 22$^{nd}$). On the otherhand, the widths of the other two curves (C2 and C3), do show a small amplitude anti-correlation trend with the solar cycles (as the variation in this case is $\sim$1 $\mathrm{Mm}$, thus the anti-correlation behaviour appear small in the axis range used in the plot). Thus, the anti-correlated behaviour of the QR scale with the solar cycle, as found by \citet{2017arXiv170500175C}, is mainly contributed by the wings of the QR scale distributions.\\ \\

\section{Summary and Conclusion}\label{sec:summary}

  In this paper, using the Kodaikanal digitized C\MakeLowercase{a}  $\scriptsize{{\textrm{II}}}$ K data, we investigated the long-term (cycle-wise) behaviour of the supergranule scales over multiple solar cycles. Below, we summarize the main results from our analysis:\\

$\bullet$ We find that the supergranule mean scales, in the AR and QR, over the long period of time (9 solar cycle), show similar kind of distributions. We also find the mean scale of AR is higher than that of the QR and the difference between the two has a maximum value of $\approx$1.8 $\mathrm{Mm}$.\\

$\bullet$ Calculations of the mean scale, for each of the solar cycles, reveal an oscillating pattern for both the AR and QR. Both of these curves depict an upward moving trend which is steeper in AR as compared to QR. Also, we find that the AR mean scale is persistently higher to QR during all the cycles ($\sim$90 years) investigated in this study. Though there exist a relatively high correlation between the two, but we also find that there are some differences in terms of the location of the maxima and minima of the mean scale curves.\\

$\bullet$ We find that these mean scales are significantly correlated with the peak TSI values. This makes them a potential calibrator for historical TSI reconstructions. The peak sunspot numbers are also correlated well with the mean scales. Interestingly, the correlation coefficients in both the cases are higher in AR as compared to QR. Among them also, we obtain a higher correlations between mean scale and peak TSI as to mean scale and peak sunspot number.\\

$\bullet$ The intra-cycle variation of the scale values are also investigated using 2-D histograms. We note an in-phase variation for the AR scale with the solar cycle cycle whereas the same for the QR are anti-correlated. The spread in the AR scale distribution is found to be higher near the start of a cycle and become narrower at the end of it whereas for the QR, the width remains almost invariant.\\

In conclusion, we see that the properties of the AR and QR (mean) scales are quite different and have different relations with solar cycles. The correlation values, indicate the possibility of using the cycle-wise mean scale as a historical TSI calibrator. Future studies using data from other observatories will greatly help in constraining  these relations which can further be used by modellers for better accuracy in the model output.

\section{Acknowledgment}
{The authors would like to thank the referee for valuable suggestions which helped to improve the quality and the presentation of the paper.
We would like to thank the Kodaikanal facility of Indian Institute of Astrophysics, Bangalore, India for proving the data. The data series is now available for public use at \url{http://kso.iiap.res.in/data}. We also thank the Science \& Engineering  Research Board (SERB) for the project grant (EMR/2014/000626).}


 \bibliographystyle{apj}

\begin{thebibliography}{30}
\expandafter\ifx\csname natexlab\endcsname\relax\def\natexlab#1{#1}\fi

\bibitem[{{Berrilli} {et~al.}(1999){Berrilli}, {Ermolli}, {Florio}, \&
  {Pietropaolo}}]{1999A&A...344..965B}
{Berrilli}, F., {Ermolli}, I., {Florio}, A., \& {Pietropaolo}, E. 1999, \aap,
  344, 965

\bibitem[{{Berrilli} {et~al.}(1998){Berrilli}, {Florio}, \&
  {Ermolli}}]{1998SoPh..180...29B}
{Berrilli}, F., {Florio}, A., \& {Ermolli}, I. 1998, \solphys, 180, 29

\bibitem[{{Chatterjee} {et~al.}(2016){Chatterjee}, {Banerjee}, \&
  {Ravindra}}]{2016ApJ...827...87C}
{Chatterjee}, S., {Banerjee}, D., \& {Ravindra}, B. 2016, \apj, 827, 87

\bibitem[Chatterjee et al.(2017)]{2017arXiv170500175C} Chatterjee, S., Mandal, S., \& Banerjee, D.\ 2017, arXiv:1705.00175 


\bibitem[Coddington et al.(2016)]{2016BAMS...97.1265C} Coddington, O., Lean, J.~L., Pilewskie, P., Snow, M., \& Lindholm, D.\ 2016, Bulletin of the American Meteorological Society, 97, 1265 

\bibitem[{{Fr{\"o}hlich} \& {Lean}(1998)}]{1998GeoRL..25.4377F}
{Fr{\"o}hlich}, C., \& {Lean}, J. 1998, \grl, 25, 4377

\bibitem[{{Fr{\"o}hlich} \& {Lean}(2004)}]{2004A&ARv..12..273F}
---. 2004, \aapr, 12, 273

\bibitem[{Hagenaar {et~al.}(1997)Hagenaar, Schrijver, \&
  Title}]{0004-637X-481-2-988}
Hagenaar, H.~J., Schrijver, C.~J., \& Title, A.~M. 1997, The Astrophysical
  Journal, 481, 988

\bibitem[{{Hart}(1956)}]{1956MNRAS.116...38H}
{Hart}, A.~B. 1956, \mnras, 116, 38

\bibitem[{Hathaway(2015)}]{lrsp-2015-4}
Hathaway, D.~H. 2015, Living Reviews in Solar Physics, 12

\bibitem[{{Hudson} {et~al.}(1982){Hudson}, {Silva}, {Woodard}, \&
  {Willson}}]{1982SoPh...76..211H}
{Hudson}, H.~S., {Silva}, S., {Woodard}, M., \& {Willson}, R.~C. 1982,
  \solphys, 76, 211

\bibitem[{Kopp {et~al.}(2016)Kopp, Krivova, Wu, \& Lean}]{Kopp2016}
Kopp, G., Krivova, N., Wu, C.~J., \& Lean, J. 2016, Solar Physics, 291, 2951

\bibitem[{{Lean} {et~al.}(1995){Lean}, {Beer}, \&
  {Bradley}}]{1995GeoRL..22.3195L}
{Lean}, J., {Beer}, J., \& {Bradley}, R. 1995, \grl, 22, 3195

\bibitem[{{Leighton} {et~al.}(1962){Leighton}, {Noyes}, \&
  {Simon}}]{1962ApJ...135..474L}
{Leighton}, R.~B., {Noyes}, R.~W., \& {Simon}, G.~W. 1962, \apj, 135, 474

\bibitem[{{Li} {et~al.}(2016){Li}, {Li}, {Zhang}, \&
  {Feng}}]{2016SoPh..291.2917L}
{Li}, K.~J., {Li}, F.~Y., {Zhang}, J., \& {Feng}, W. 2016, \solphys, 291, 2917

\bibitem[{Lin {et~al.}(2003)Lin, Adiga, Olson, Guzowski, Barnes, \&
  Roysam}]{CYTO:CYTO10079}
Lin, G., Adiga, U., Olson, K., Guzowski, J.~F., Barnes, C.~A., \& Roysam, B.
  2003, Cytometry Part A, 56A, 23

\bibitem[{Lockwood \& Fr{\"o}hlich(2007)}]{lockwood2007recent}
Lockwood, M., \& Fr{\"o}hlich, C. 2007in , The Royal Society, 2447--2460

\bibitem[{McIntosh {et~al.}(2011)McIntosh, Leamon, Hock, Rast, \&
  Ulrich}]{2041-8205-730-1-L3}
McIntosh, S.~W., Leamon, R.~J., Hock, R.~A., Rast, M.~P., \& Ulrich, R.~K.
  2011, The Astrophysical Journal Letters, 730, L3

\bibitem[{{Meunier} {et~al.}(2008){Meunier}, {Roudier}, \&
  {Rieutord}}]{2008A&A...488.1109M}
{Meunier}, N., {Roudier}, T., \& {Rieutord}, M. 2008, \aap, 488, 1109

\bibitem[{{Muenzer} {et~al.}(1989){Muenzer}, {Schroeter}, {Woehl}, \&
  {Hanslmeier}}]{1989A&A...213..431M}
{Muenzer}, H., {Schroeter}, E.~H., {Woehl}, H., \& {Hanslmeier}, A. 1989, \aap,
  213, 431

\bibitem[{Paniveni {et~al.}(2010)Paniveni, Krishan, Singh, \&
  Srikanth}]{Paniveni11022010}
Paniveni, U., Krishan, V., Singh, J., \& Srikanth, R. 2010, Monthly Notices of
  the Royal Astronomical Society, 402, 424

\bibitem[{{Plaskett}(1916)}]{1916ApJ....43..145P}
{Plaskett}, H.~H. 1916, \apj, 43, 145

\bibitem[{{Priyal} {et~al.}(2014){Priyal}, {Singh}, {Ravindra}, {Priya}, \&
  {Amareswari}}]{2014SoPh..289..137P}
{Priyal}, M., {Singh}, J., {Ravindra}, B., {Priya}, T.~G., \& {Amareswari}, K.
  2014, \solphys, 289, 137

\bibitem[{Rast(2003)}]{0004-637X-597-2-1200}
Rast, M.~P. 2003, The Astrophysical Journal, 597, 1200

\bibitem[{Rieutord \& Rincon(2010)}]{Rieutord2010}
Rieutord, M., \& Rincon, F. 2010, Living Reviews in Solar Physics, 7, 2

\bibitem[{{Simon} \& {Leighton}(1964)}]{1964ApJ...140.1120S}
{Simon}, G.~W., \& {Leighton}, R.~B. 1964, \apj, 140, 1120

\bibitem[{Srikanth {et~al.}(2000)Srikanth, Singh, \&
  Raju}]{0004-637X-534-2-1008}
Srikanth, R., Singh, J., \& Raju, K.~P. 2000, The Astrophysical Journal, 534,
  1008

\bibitem[{{S{\'y}kora}(1970)}]{1970SoPh...13..292S}
{S{\'y}kora}, J. 1970, \solphys, 13, 292

\bibitem[{{Willson}(1997)}]{1997Sci...277.1963W}
{Willson}, R.~C. 1997, Science, 277, 1963

\bibitem[{{Willson} {et~al.}(1981){Willson}, {Gulkis}, {Janssen}, {Hudson}, \&
  {Chapman}}]{1981Sci...211..700W}
{Willson}, R.~C., {Gulkis}, S., {Janssen}, M., {Hudson}, H.~S., \& {Chapman},
  G.~A. 1981, Science, 211, 700

\end{thebibliography}

\end{document}